\documentclass[12pt]{article}
\usepackage{graphicx}
\newcommand{\beq}{\begin{equation}}
\newcommand{\eeq}{\end{equation}}
\newcommand{\bea}{\begin{eqnarray}}
\newcommand{\eea}{\end{eqnarray}}

\def\lsim{~\,\makebox(1,1){$\stackrel{<}{\widetilde{}}$}\,~}

\begin{document}
\begin{titlepage}
\begin{flushleft}
       \hfill                      {\tt hep-th/0501xxx}\\
       \hfill                       FIT HE - 05-01 \\
       \hfill                       KYUSHU-HET-80 \\
\end{flushleft}
\vspace*{3mm}
\begin{center}
{\bf\LARGE Flavor quark at high temperature \\
from a holographic model}

\vspace*{5mm}
\vspace*{12mm}
{\large Kazuo Ghoroku\footnote[2]{\tt gouroku@dontaku.fit.ac.jp},
Tomohiko Sakaguchi\footnote[3]{\tt tomohiko@higgs.phys.kyushu-u.ac.jp}
Nobuhiro Uekusa\footnote[3]{\tt uekusa@higgs.phys.kyushu-u.ac.jp}
and Masanobu Yahiro\footnote[3]{\tt yahiro2scp@mbox.nc.kyushu-u.ac.jp}
}\\
\vspace*{2mm}

\vspace*{2mm}

\vspace*{4mm}
{\large ${}^{\dagger}$Fukuoka Institute of Technology, Wajiro, 
Higashi-ku}\\
{\large Fukuoka 811-0295, Japan\\}
\vspace*{4mm}
{\large ${}^{\ddagger}$Department of Physics, Kyushu University, Hakozaki,
Higashi-ku}\\
{\large Fukuoka 812-8581, Japan\\}

\vspace*{10mm}
\end{center}

\begin{abstract}

Gauge theory with light flavor quark is studied by embedding a D7 brane 
in a deconfinement phase background newly constructed. We find a phase
transition by observing a jump of the vacuum expectation value of quark 
bilinear and also of the derivative of D7 energy at a critical temperature.
For the model considered here, we also study quark-antiquark potential 
to see some possible quark-bound states and other physical quantities
in the deconfinement phase. 

\end{abstract}
\end{titlepage}

\section{Introduction}

It is still a challenging problem to make clear the gauge/gravity
correspondence from superstring theory \cite{MGW}. 
In particular, we expect that this correspondence 
is applicable to QCD by deforming the anti-de Sitter space-time 
(AdS) into an appropriate non-conformal form.

Recently an idea to add light flavor quarks 
has been 
proposed by Karch and Katz~\cite{KK} for D3-D7 brane system in the
AdS$_5\times S^5$ background. After that,
several authors have extended this idea to various 10d gravity backgrounds
corresponding to the various gauge duals, and they have examined the
meson spectra and chiral symmetry breaking in the context of the holography
~\cite{KMMW,KMMW2,Bab,ES,SS,NPR,
GY,sakaisugimoto,BY}. 
There would be many directions to extend this idea. An interesting
direction would be the analysis at finite temperature which is given in 
~\cite{KMMW2} for the D4-D6 model. However many things are left to be
examined for the case of finite temperature.

Here we give such analyses in the background which is obtained as the 
extended solution to the finite temperature of the one given in \cite{GY}.
The background given here corresponds to the Yang-Mills theory in the
deconfining, high-temperature phase. The D7 brane is embedded in this 
background, and we could observe a gap of
the vacuum expectation value of quark bilinear and also of
the derivative of D7 energy with respect to the temperature.
This implies a phase transition in the gauge theory at some temperature, 
and we discuss this point. And the problem related to the chiral symmetry is
also discussed.

\vspace{.3cm}
Through the estimation of the
Wilson-Polyakov loop, we obtain a static
quark-antiquark potential at finite-temperature, 
which is very similar to the one given by Rey,
Theisen and Yee for the infinitely heavy quarks in the 
AdS$_5\times S^5$ background~\cite{RTY}. We also estimate the dynamical 
quark mass, and we discuss these results
by comparing them with numerical results given in the recent lattice gauge
simulations.

Especially, the potential obtained here
implies that some meson states would remain until the temperature exceeds 
a critical value~\cite{Jackiw,Malik}, 
which is estimated here. A similar phenomenon is also
seen for D5 baryon state at finite temperature. We discuss on these points.

\vspace{.2cm}
In section 2, we give the setting of our model, and a phase
transition is pointed out by embedding the D7 brane. 
In section 3, the quark-antiquark potential and the dynamical quark mass 
are studied through the Wilson-Polyakov loop estimations. 
In section 4, possible bound states for meson and baryon are
discussed, and we also estimate the screening mass.
The summary is given in the final section.

\section{Background geometry}

We solve the equations of motion for 10d IIB model under the Freund-Rubin
ansatz for selfdual five form field strength, 
$F_{\mu_1\cdots\mu_5}=-\sqrt{\Lambda}/2~\epsilon_{\mu_1\cdots\mu_5}$ 
\cite{KS2,LT},
and the following solution is obtained.
The solution is written 
in the string frame and taking $g_s=1$, as follows,
$$ 
ds^2_{10}=G_{MN}dX^{M}dX^{N} ~~~~~~~~~~~~~~\hspace{6.5cm}
$$ 
\beq
=e^{\Phi/2}
\left\{
{r^2 \over R^2}\left(-f^2(r)dt^2+(dx^i)^2\right)+
{1\over f^2(r)}\frac{R^2}{r^2} dr^2+R^2 d\Omega_5^2 \right\} \ , 
\label{finite-T-sol}
\eeq 
\beq
e^\Phi= \left( 1+\frac{q}{r_T^4}\log({1\over 1-(r_T/r)^4}) \right) \ , \quad \chi=-e^{-\Phi}+\chi_0 \ ,
\label{dilaton}
\eeq
\beq
  f(r)=\sqrt{1-({r_T\over r})^4} , \label{tempe}
\eeq
where $M,~N=0\sim 9$, 
$R^4=4 \pi N$ and $q$ is a constant which represent the vev of gauge fields
condensate~\cite{GY}. $\Phi$ and $\chi$
denote the dilaton and the axion respectively.
And other field configurations are set to be zero here.
The temperature $T$ is related to the parameter $r_T$ as $r_T=\pi R^2 T$.

\vspace{.3cm}
In the background given above, we study the dynamical properties of
flavor quarks which are introduced as the strings connecting the stacked
D3 branes and a newly embedded D7 brane as a probe. The D7 brane is 
embedded as follows. The six dimensional part of the above metric
is rewritten as
\beq
 {1\over f^2(r)}\frac{R^2}{r^2} dr^2+R^2 d\Omega_5^2
 =\frac{R^2}{U^2}\left(d\rho^2+\rho^2d\Omega_3^2+(dX^8)^2+(dX^9)^2
\right)\ ,
\eeq
\beq
  U(r)=\exp\left( \int{dr\over r\sqrt{1-(r_T/r)^4}} \right)
     =r\sqrt{{1+f(r)\over 2}}.
\eeq
Here $U$ is normalized as $U=r$ for $r_T=0$, and $U^2=\rho^2+(X^8)^2+(X^9)^2$.
Then we obtain the induced metric for D7 brane,
$$ 
ds^2_8=e^{\Phi/2}
\left\{
{r^2 \over R^2}\left(-f^2(r)dt^2+(dx^i)^2\right)+\right.
\hspace{3cm}
$$
\beq
\left.\frac{R^2}{U^2}\left((1+(\partial_{\rho}w^8)^2
+(\partial_{\rho}w^9)^2)d\rho^2+\rho^2d\Omega_3^2\right)
 \right\} \ , 
\label{D7-metric}
\eeq
where $w^8(\rho)$ and $w^9(\rho)$ are the scalars which determine
the position of D7 brane. They are solved under the ansatz that
they depend on only $\rho$. Further we can set $w^9=0$ and $w^8=w(\rho)$
without loss of generality due to the rotational invariance in
$X^8-X^9$ plane. 

\vspace{.3cm}
The brane action for the D7-probe is given as
\beq
S_{\rm D7}= -\tau_7 \int d^8\xi \left(e^{-\Phi} \sqrt{\cal G} 
      +{1\over 8!}\epsilon^{i_1\cdots i_8}A_{i_1\cdots i_8}\right) \ ,
\label{D7-action}
\eeq
where ${\cal G}=-{\rm det}({\cal G}_{i,j})$, $i,~j=0\sim 7$. 
${\cal G}_{ij}= \partial_{\xi^i} X^M\partial_{\xi^j} X^N G_{MN}$
and $\tau_7$ represent the induced metric and 
the tension of D7 brane respectively. Here we consider the case
of zero $U(1)$ gauge field on the brane, 
but we notice that the eight form potential $A_{i_1\cdots i_8}$, 
which is Hodge dual to the axion, couples to the 
D7 brane minimally. We obtain the eight form potential $A_{(8)}$
as $F_{(9)}=dA_{(8)}$ in terms of the Hodge dual field strength $F_{(9)}$ 
\cite{GGP}.
By taking the canonical gauge, we arrive at the following D7 brane
action,
\beq
S_{\rm D7} =-\tau_7~\int d^8\xi  \sqrt{\epsilon_3}\rho^3
\left(({r\over U})^4{f}
   e^{\Phi}\sqrt{ 1 + (w')^2 }+C_8 \right)
\ ,
\label{D7-action-2}
\eeq
\beq
 C_8=-{q\over U^4}. \label{A8}
\eeq
Here we notice that $C_8\to -q/r^4(=1-e^{\Phi})$ for $r_T\to 0$, and 
this is consistent with the previous result at $T=0$.
We solve the equation of motion for $w(\rho)$, 
\bea
   (w-\rho w')
   \left[2(1-f)^2-{q\over r^4}e^{-\Phi}\right]
   \!\!\!&+&\!\!\!4w\sqrt{1+(w')^2}{q\over r^4}e^{-\Phi}
\nonumber\\
   &-&\!\!\!  U^2 f\left[3{w'\over \rho}+ {w''\over 1+(w')^2}\right]
   =0 ,
  \label{qeq}
\eea
and find a suitable embedding configuration used in the analysis given
here.
Here we notice some points with respect to the above action. 
Firstly we expect that the solutions for $w$ at $r_T=0$ would be
smoothly connected to the one of finite $r_T$ although
the horizon appears in the background for finite $T$. However
we find a phase transition when the end point
$w(0)$ of the solution jumps from $w(0)$ to $w(\rho_0)$ 
where $\rho_0$ is a point on the horizon.
We discuss this point through the embedding solutions.

\vspace{.3cm}
The solution $w$ for large $\rho$ has the asymptotic form
\beq
   w(\rho) \sim m_q+{c\over \rho^2} ,  \label{asym}
\eeq
where $m_q$ and $c$ are interpreted from the gauge/gravity 
correspondence as
the current quark mass and the chiral condensate,
respectively. We find that $w(\rho)=0$ ($m_q=0$ and $c=0$) is always the 
solution of (\ref{qeq}), and any other solution of
nonzero $m_q$ leads necessarily to nonzero and negative $c < 0$. 
In other words, the chiral symmetry is preserved only for the solution
$w(\rho)=0$. We notice however that the sign of nonzero $c$ is opposite to the
case of spontaneous chiral symmetry breaking. This is because of the
attractive force between D3 and D7 branes in the present case.

\begin{figure}[htbp]
\vspace{.3cm}
\begin{center}
  \includegraphics[width=11.5cm]{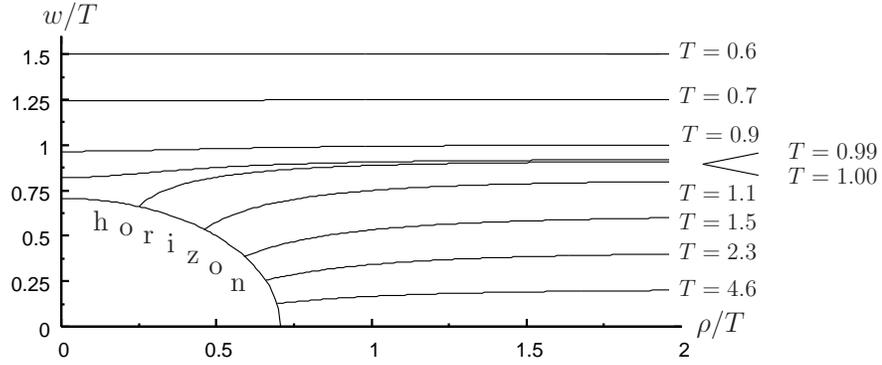}
\caption{Embedding solutions for $q=0$. 
The solutions are drawn for several temperatures, where
$m_q=0.91$ and $\pi R^2=1$.
 \label{wq0fig}}
\end{center}
\end{figure}

The temperature dependence of the solution is shown in the 
Fig.\ref{wq0fig} for $q=0$. We notice that this result is equivalent
to the one given in Ref.~\cite{Bab} when $T$ is replaced by $m_q$.
This is because of the same form of equations for $w$ and its independence
from the rescaling of all mass dimensional parameters. Actually it is 
possible to replace $T$ by $m_q$ by an appropriate normalization.

We find a jump of the solution near $T=1$, and we expect this as 
some kind of phase transition. 
We are already considering in the deconfining phase, so we suppose that there
is no hadronic bound state in this phase. However there would be a possible
region, as shown below,
of the temperature where some hadronic states are still remaining. So,
we might expect that a transition from a phase with hadronic states 
to a phase without any hadronic state would occur at
some temperature. Namely, all
the remaining bound states in the high temperature phase disappear
above this critical temperature.
More on this point, we discuss below and in the following sections.
\begin{figure}[htbp]
\vspace{.3cm}
\begin{center}
  \includegraphics[width=11.5cm]{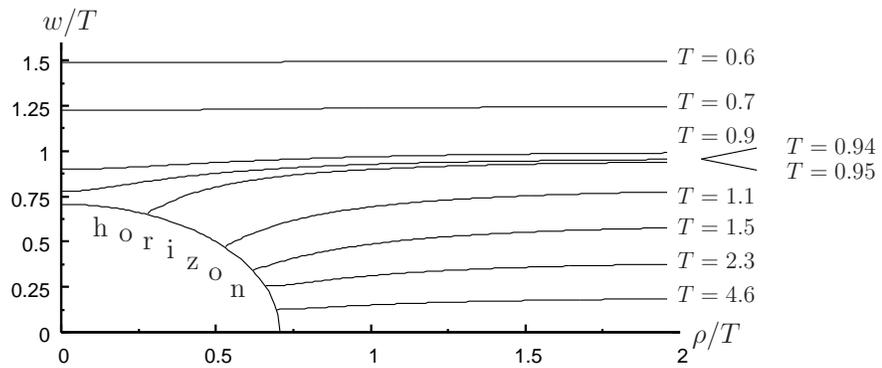}
\caption{Embedding solutions for $q/T^4=0.1$ and $m_q=0.91$.
The way of the embedding changes at $T = 0.94\sim 0.95$ in unit of 
$\pi R^2=1$.
 \label{wq1fig}}
\end{center}
\end{figure}

\vspace{.3cm}
For $q\neq 0$, from the equation (\ref{qeq}) we find embedding solutions 
which are shown in Figure~\ref{wq1fig}.
As expected, the gauge field condensate $q$ affects
the critical temperature, 
and it moves to smaller value than that of the case of $q=0$.
This implies that 
the critical temperature decreases
in the presence of the gauge field condensation.
This is understood as follows. For nonzero $q$, the force becomes small
so it would need lower temperature to make the quarks being free
than that of the case of $q=0$.

\begin{figure}[htbp]
\vspace{.3cm}
\begin{center}
  \includegraphics[width=9cm]{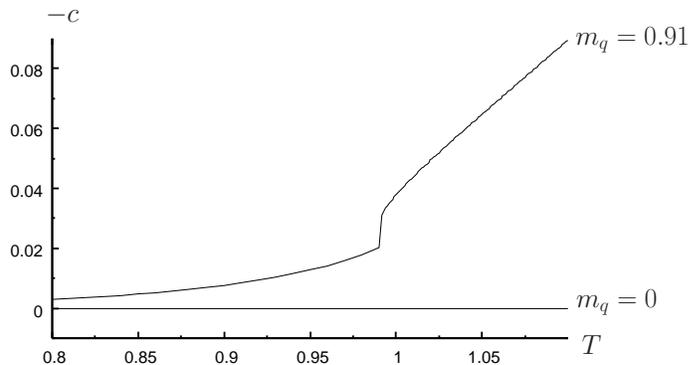}
\caption{The temperature dependence of the chiral condensate:
for $m_q = 0.91$ and $\pi R^2=1$.
 \label{ctfig}}
\end{center}
\end{figure}

\vspace{.3cm}
Now we would like to investigate the chiral condensate $c$ and 
the energy of D7-brane for the embedding solution.
The shape of the solution at high temperature would be determined 
mainly by the
factor $f(r)$ and the effect from finite $q$ would be miner. So we
study the high temperature solution of $w$ at $q=0$ for simplicity.
The chiral condensate depends only on temperature
when $m_q$ and $R$ are fixed.
The absolute value of $c$ is large at high temperature 
where the internal coordinates has an endpoint on the horizon.
At low temperature, its value becomes small.
This behavior is shown in Fig.~\ref{ctfig}.
From this figure, it is seen
that a phase transition occurs at $T\sim 1$ for $m_q = 0.91$.
This is consistent with the phase transition which was found by
changing the value of $m_q$ with fixed $T$~\cite{Bab}.

\begin{figure}[htbp]
\vspace{.3cm}
\begin{center}
  \includegraphics[width=12cm]{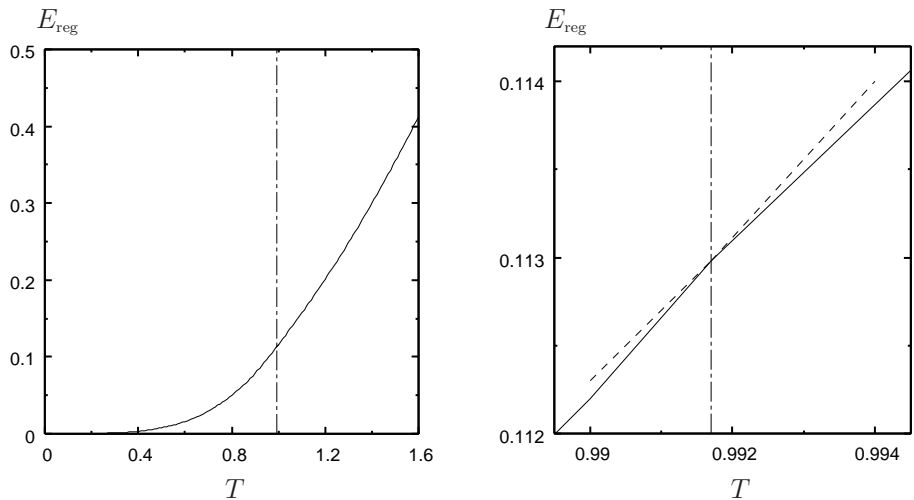}
\caption{The temperature dependence of the regularized energy:
for $m_q = 0.91$ and $\pi R^2=1$.
The vertical line denotes $T=T_{\textrm{\scriptsize fund}}$.
The right figure shows the neighborhood of the transition point.
The dashed lines stand for the slops at the transition point.
 \label{etfig}}
\end{center}
\end{figure}

\vspace{.3cm}
We turn to temperature dependence of the D7-brane energy.
By substituting the background (\ref{finite-T-sol}) $\sim$ (\ref{tempe})
into the D7-brane action  (\ref{D7-action-2}),
the D7-brane energy for $q=0$ is written as
\beq
  E_{\textrm{\scriptsize D7}}
  =\int_{\rho_{\textrm{\scriptsize min}}}^{\infty} d\rho\,\, \rho^3
  \left(1-{r_T^8\over 16 U^8}\right)
  \sqrt{1+(w')^2} , \label{d7e}
\eeq
which is scaled by $\tau_7 \sqrt{\epsilon_3}$.
The lower bound
$\rho_{\textrm{\scriptsize min}}$ is either zero, or, a point on the
horizon which the-brane meets.
The integral (\ref{d7e}) diverges.
We regularize it by subtracting the D7-brane energy for $m_q=0$
in similar to the analysis given for D4- and D6-branes
in Ref.~\cite{KMMW}\footnote{Note that
the chiral condensate and the regularized energy in Ref.~\cite{KMMW} are 
rescaled by the temperature.}.
The regularized energy is
\bea
  E_{\textrm{\scriptsize reg}}&=&
  E_{\textrm{\scriptsize D7}}(m_q)-E_{\textrm{\scriptsize D7}}(0)
\nonumber\\  
&=&\int_{\rho_{\textrm{\scriptsize min}}}^{\rho_{\textrm{\scriptsize
  match}}}
  d\rho\,\, \rho^3
  \left(1-{r_T^8\over 16 U^8}\right)
  \sqrt{1+(w')^2}
\nonumber\\ 
&&\qquad\qquad
  -\int_{\rho_H}^{\rho_{\textrm{\scriptsize
  match}}}
  d\rho\,\, 
  \left(\rho^3-{r_T^8\over 16 \rho^5}\right)
  +{c^2\over \rho_{\textrm{\scriptsize match}}^2} .
\eea
where $\rho_H=r_T/\sqrt{2}$ and $\rho_{\textrm{\scriptsize match}}$ is 
the point where
we match numerically $w$ to the asymptotic solution (\ref{asym}).
The last term is corrections from the integration for
$\rho >\rho_{\textrm{\scriptsize match}}$ up to
${\cal O}(\rho_{\textrm{\scriptsize match}}^{-4})$.
The regularized energy increases monotonically
with temperature.
For fixed $m_q$, the slope $dE_{\textrm{\scriptsize reg}}/dT$
has a discontinuous jump at $T=T_{\rm fund}$. 
Fig.~\ref{etfig} shows the temperature dependence
of the regularized energy.
It is clear that
the lowest D7-brane energy is obtained for the case where 
the current quark mass is zero.
For $m_q\neq 0$,
the energy depends on whether the end point of $w$ is
on the horizon or not.
At high temperature side, $T>T_{\textrm{\scriptsize fund}}$,
the energy of the solution which meet the horizon
becomes lower than the one of the other type solution.

As we will show, the solution attached to the horizon 
leads to vanishing of dynamical quark mass and then of the quark-antiquark 
potential.
Then the change of phase at $T=T_{\textrm{\scriptsize fund}}$ will be 
regarded as the phase transition from the phase with surviving hadronic states 
to the free-quark phase. This point will be discussed more in the followings.

\vspace{.3cm}

\section{Quark-antiquark potential}

We study a gravity description of 
quark-antiquark potentials in detail.
Before performing concrete calculations, 
we review how quark-antiquark potentials are described
in the context of the gauge/gravity correspondence. 
The point relevant to the present purpose is in the following.

We consider the Wilson-Polyakov loop in $SU(N)$ gauge theory:
\beq
   W={1\over N} \textrm{Tr} P e^{i\int A_0 dt} .
\eeq
The quark-antiquark potential $V_{q\bar{q}}$ is derived from
the expectation value of a parallel Wilson-Polyakov loop: 
\beq
   \langle W\rangle \sim e^{-V_{q\bar{q}}\int dt} . \label{wpl}
\eeq
On the other hand, the dual gravity suggests
that the expectation value is represented as
\beq
    \langle W\rangle  \sim e^{-S} , \label{wstr}
\eeq
in terms of the Nambu-Goto action 
\beq
   S=- \frac{1}{2 \pi \alpha'} 
\int d\tau d\sigma \sqrt{-\textrm{det}\, h_{ab}} , 
\eeq
with the induced metric
\beq
    h_{ab}=G_{\mu\nu}\partial_a X^{\mu}\partial_b X^{\nu} ,
\eeq   
where the string coordinate is $X^{\mu}(\tau,\sigma)$ and
the string world-sheet is parameterized by $\sigma$, $\tau$.
From the equations (\ref{wpl}) and (\ref{wstr}),
the quark-antiquark potential can be calculated
by setting various configurations of string coordinates and background
geometries.
In the following analysis, 
we investigate  quark-antiquark potentials
by considering static string configurations.
\subsection{Gauge-field condensate model}

We examine quark-antiquark potentials
in  the background presented here. 
To study possible static string configurations 
of a pair of quark  and anti-quark,
we choose $X^0=t=\tau$ and decompose 
the other nine string coordinates into components 
parallel and perpendicular to the D3-branes:
\beq
 \mathbf{X} = (\mathbf{X}_{||}, r, r \Omega_5) .
\eeq
The Nambu-Goto Lagrangian in the background (\ref{finite-T-sol}) 
becomes
\beq
   L_{\textrm{\scriptsize NG}}=-{1 \over 2 \pi \alpha'}\int d\sigma ~e^{\Phi/2}
   \sqrt{r'{}^2+r^2f(r)^2\Omega_5{}'{}^2
        +\left({r\over R}\right)^4 f(r)^2 \mathbf{X}_{||}{}'{}^2} ,
 \label{ng}
\eeq
where the prime denotes a derivative with respect to $\sigma$.
The test string has two possible configurations:
(i) a pair of parallel strings, which connect horizon and the D7 brane,
and (ii) a U-shaped string whose two end-points are on the D7 brane.

\vspace{.3cm}
We firstly consider the configuration (i) of parallel two strings,
which have no correlation each other.
The total energy is then two times of one dynamical quark mass,
$\tilde{m}_q$. As mentioned
above, it is given by a string configuration which stretches 
between the horizon $r_T$ and the maximum $r_{\textrm{\scriptsize max}}$,
so we can take as
\beq
   r=\sigma,~~~ \mathbf{X}_{||}=\textrm{constant},~~~
   \Omega_5=\textrm{constant}. 
\label{para}
\eeq
Then $\tilde{m}_q$ is obtained by substituting 
(\ref{para}) into (\ref{ng}) as follows,
\beq
   E=
   {1\over \pi \alpha'}\int_{r_T}^{r_{\textrm{\scriptsize max}}}
     dr ~e^{\Phi/2} = 2\tilde{m}_q \ . \label{dynamicalmass}
\eeq

\begin{figure}[htbp]
\vspace{.3cm}
\begin{center}
  \includegraphics[width=10cm]{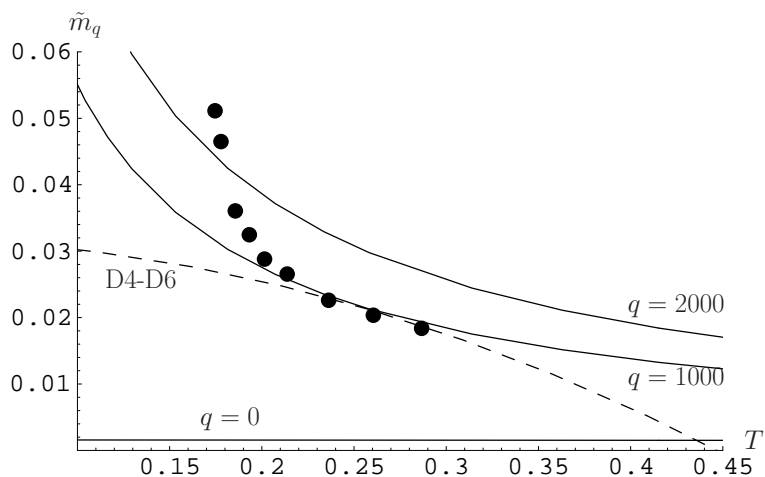}
\caption{
 $\tilde{m}_q$ 
are shown for $R=1/\sqrt{\pi} (\mbox{GeV}^{-1})$, 
$r_{\textrm{\scriptsize max}}=10 (\mbox{GeV}^{-1})$ 
and $\alpha'=10^3 (\mbox{GeV}^{-2})$. 
The three solid curves are corresponding to the case of 
$q=0,~10^3$ and $2\times 10^3 (\mbox{GeV}^{-4})$, respectively. 
The dashed curve represents the result for 
the D4-D6 model~(\ref{dynamicalmass46}) with 
$R =10 (3/{4\pi})^{2/3}$, $r_{\textrm{\scriptsize max}}=200$ 
and $\alpha'=10^3$. 
The points represent the lattice data ~\cite{lattice}. }
 \label{mqT}
\end{center}
\end{figure}

\vspace{.3cm}
The temperature dependence of the dynamical mass $\tilde{m}_q$ 
is shown in Fig.~\ref{mqT}.
%

Generally, $r_{\textrm{\scriptsize max}}$ depends on temperature.
However, when temperature is low, its change is very small.
Therefore we set approximately constant $r_{\textrm{\scriptsize max}}$.
The points are quoted 
from the lattice data of Fig.~5 in reference~\cite{lattice}:
we regard the asymptotic values of the heavy quark free energy
as the sum of two dynamical quark masses. 
\footnote{We convert the data by use of $T_c \cong 0.173$ GeV and
$T_c/\sqrt{\sigma} \cong 0.425$, where $\sigma$ is a string tension.}
The behavior of our result is resemble to the lattice behavior
because of the concave shape.
Especially,   
putting $R = 1/\sqrt{\pi} (\mbox{GeV}^{-1})$, 
$q=10^3 (\mbox{GeV}^{-4})$, 
$r_{\scriptsize{max}}=10 (\mbox{GeV}^{-1})$ 
and $\alpha' = 10^3 (\mbox{GeV}^{-2})$, 
the dynamical mass~(\ref{dynamicalmass}) corresponds to 
the lattice results at least 
for the region $0.2 \lsim T \lsim 0.3 (\mbox{GeV})$.
Although $\alpha'$ has to become small,
we can fit the result of the model to the lattice data
only at a large value i.e. $\alpha' = 10^3 (\mbox{GeV}^{-2})$.  
This situation is an open problem here.

For $q=0$, temperature dependence of $\tilde{m}_q$ is not seen, but
it largely affected by $T$ for $q\neq 0$. And, at any point of $T$, 
$\tilde{m}_q$ increases with $q$. We should notice that $\tilde{m}_q$
disappears when the temperature exceeds $T_{\textrm{\scriptsize fund}}$,
the D3-brane is included in the D7-brane.

\vspace{.5cm}
We now turn to the U-shaped configuration,
\beq
    \mathbf{X}_{||}=(\sigma,0,0),~~~
    \Omega_5=\textrm{constant} .
    \label{ushape}
\eeq
The equation of motion derived from the Lagrangian (\ref{ng}) with
the configuration (\ref{ushape})
are solved by
\beq
     e^{\Phi/2}{1\over \sqrt{(r/R)^4 f(r)^2+(dr/d\sigma)^2}}
    \left({r\over R}\right)^4 f(r)^2= \textrm{constant} .
\eeq
The midpoint $r_0$ of the string is determined by $dr/d\sigma|_{r=r_0}=0$.
Then the distance and the total energy of the quark and anti-quark
are given by
\bea
  &&  L=2R^2 \int_{r_0}^{r_{\textrm{\scriptsize max}}} dr~
      {1\over r^2 f(r)
        \sqrt{e^{\Phi(r)}r^4 f(r)^2 /
          \left(e^{\Phi(r_0)}r_0^4f(r_0)^2\right)-1}} ,
\\
  && E=
   {2\over \pi \alpha'} \int_{r_0}^{r_{\textrm{\scriptsize max}}}
   {e^{\Phi(r)/2}\over 
     \sqrt{1-e^{\Phi(r_0)}r_0^4 f(r_0)^2/
             \left(e^{\Phi(r)}r^4 f(r)^2\right)}} .
\eea

\begin{figure}[htbp]
\vspace{.3cm}
\begin{center}
$
  \includegraphics[width=7.5cm]{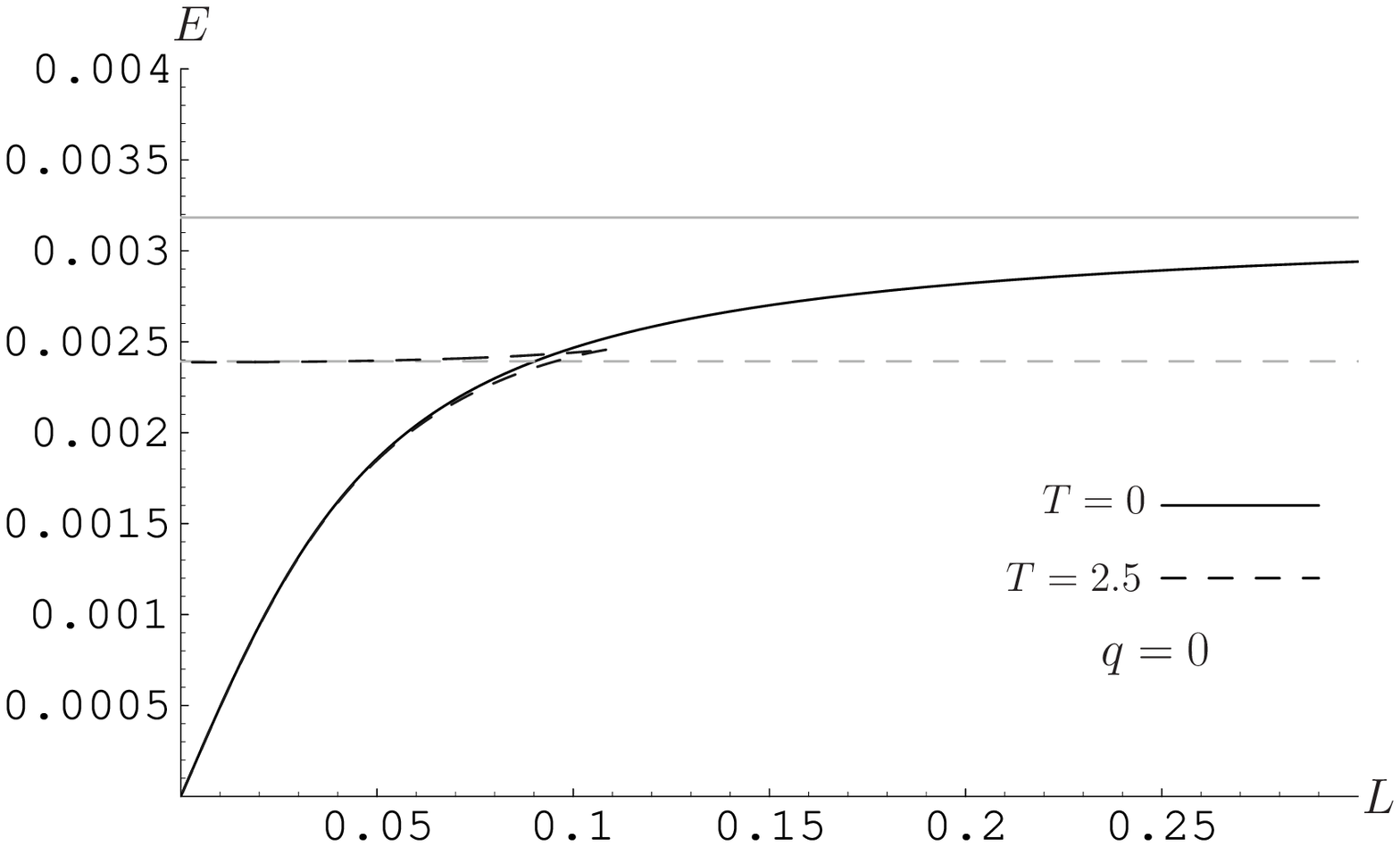} \
 \includegraphics[width=7.5cm]{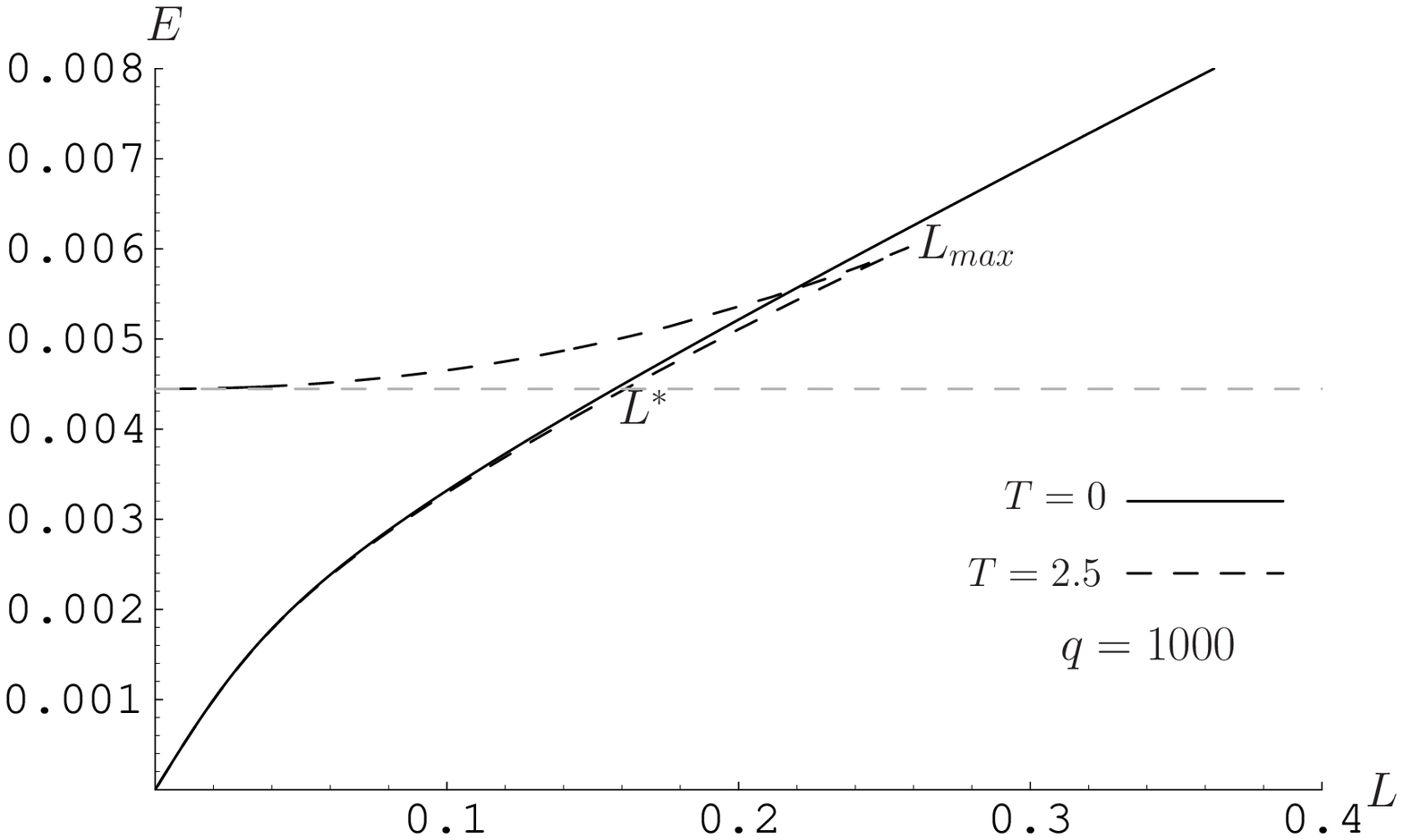}
$
\caption{Plots of $E$ vs $L$ at $q=0$ and $q=10^3(\mbox{GeV}^{-4})$
for $R = 1/ \sqrt{\pi} (\mbox{GeV}^{-1})$, 
$r_{\textrm{\scriptsize max}}=10 (\mbox{GeV}^{-1})$ 
and $\alpha'=10^3 (\mbox{GeV}^{-2})$.
The solid and dashed curves represent the case of 
$T=0 $ and $T=2.5(\mbox{GeV})$, respectively. 
The vertical solid and dashed lines represent the energy of two 
parallel straight strings.
\label{el}}
\end{center}
\end{figure}
Figure~\ref{el} shows the dependence of the energy $E$
on the distance $L$ at some selected temperatures $T$ and $q$.
The results at $q=0$ (left Fig.~\ref{el}) are consistent with the one given
in reference~\cite{RTY}, where infinitely heavy quarks are considered.
However, we consider the quark with a
light mass, not the heavy quark, we then need not to consider the energy
difference between the U-shape string and a pair of strings 
as in ~\cite{RTY}. 

\vspace{.3cm}
For the case of finite $q$ (the right Fig.~\ref{el}), 
we can see the linear rising
potential for $T=0$, and it shows the confinement of 
quark and antiquark. 
On the other hand,
for finite temperature $T = 2.5$, the qualitative behavior coincides with the
one of $q=0$. Namely, 
$E$ increases with $L$ along the curve of $T=0$ but the potential disappears
at $L=L_{\scriptsize{max}}$, which depends on the temperature.
More important fact is that $E$ exceeds 
the energy of the two straight strings
configuration at $L = L^* <L_{\scriptsize{max}}$. 
When $L \geq L^*$, the straight strings configuration has a lower energy
than the U-shaped string configuration.
As the two straight strings have no interaction energy,
this shows that the quark-antiquark potential vanishes for $L \geq L^*$.
So there will be no physical meaning the potential obtained in the
region of $L^*<L<L_{\scriptsize{max}}$      
This characteristic behavior is qualitatively in agreement with 
the suggestions given by lattice simulations~\cite{lattice}.


\subsection{D4-D6 model}
Next, we calculate the quark-antiquark
potential by using the D4-D6
model~\cite{KMMW2}.
The type IIA supergravity background dual to $N_c$ D4-branes
compactified in a circle with anti-periodic boundary conditions for the 
fermions at high temperatures takes the form  
\begin{eqnarray}
 d s^2 \!\!\!\! &=& \!\!\!\! 
\left( \frac{r}{R} \right)^{3/2} \left(- \tilde{f}(r) d t^2 
+ \sum_{i=1}^3 d x^i dx^i + d \tau^2  \right) 
+ \left( \frac{R}{r} \right)^{3/2} \frac{dr^2}{\tilde{f}(r)}
+ R^{3/2} r^{1/2} d \Omega_4^2 \ , \\
&& e^{\Phi} = \left( \frac{r}{R} \right)^{3/4} \ , \,\,\, \,\,\, 
\tilde{f}(r) = 1-\frac{r_T^3}{r^3} \ , \label{D4D6background}    
\end{eqnarray} 
at high temperature.
The coordinates $(t, x^1, x^2, x^3)$ parametrize the four
directions along the D4-branes 
and time-coordinate $t$ is compactified with the period $1/T$.
The coordinate $\tau$ parametrizes the circular 4th-direction on which the
branes are compactified. 
$d \Omega_4^2$ is the $SO(5)$-invariant line element. 
$r$ has dimensions of length and is regarded as a radial coordinate 
in the 56789-directions transverse to the D4-branes.
Since we wish to avoid conical singularities at $r=r_T$,
the boundary condition fixes the metric parameter as 
\begin{eqnarray}
 r_T = \left(\frac{4 \pi T}{3}  \right)^2 R^3 \ . \label{D4D6metricparameter}
\end{eqnarray}
We take $X^0 = t$ and decompose the nine spatial
embedding coordinates as follows:
\begin{eqnarray}
 \mathbf{X} = (\mathbf{X}_{||}, T, r, r \Omega_4)
\end{eqnarray}
In this case,
The Nambu-Goto Lagrangian in a static configuration
becomes 
\begin{eqnarray}
 L_{\textrm{\scriptsize NG}} = - \frac{1}{2 \pi \alpha'} \int d \sigma 
\sqrt{ r^{'2} + \tilde{f}(r) r^2 \Omega_4^{'2} + \left( \frac{r}{R} \right)^3
\tilde{f}(r) ( \mathbf{X}^{'2} + T^{'2}) } \ .
\end{eqnarray}

\vspace{.5cm}
As in previous section, 
we set the embedding coordinates for a pair of straight strings
which are stretched between D4-branes and D6-brane as follows:
\begin{eqnarray}
r = \sigma \ , \,\,\, \,\,\, 
\mathbf{X}_{||}=\mbox{constant} \ , \,\,\, \,\,\, 
T=\mbox{constant} \ , \,\,\, \,\,\, 
\Omega_4=\mbox{constant} \ ,
\end{eqnarray}  
so that we obtain the total energy of the quark-antiquark pair
\begin{eqnarray}
 E = \frac{1}{\pi \alpha'} (r_{\textrm{\scriptsize max}}-r_T) 
= 2 \tilde{m}_q \ .
\label{dynamicalmass46}
\end{eqnarray}
Because the position of the horizon $r_T$ is proportional to 
the square of temperature, 
the temperature dependence of the dynamical quark mass $\tilde{m}_q$ 
has a convex form (see Fig.\ref{mqT}).
This form is considerably different from the tendency of the lattice result
and also from the result our gauge-field condensate model. 
Also, the dynamical mass becomes zero before the horizon approaches to the
D6-brane and this tendency is different from the gauge-field condensate model. 

\begin{figure}[htbp]
\vspace{.3cm}
\begin{center}
  \includegraphics[width=9cm]{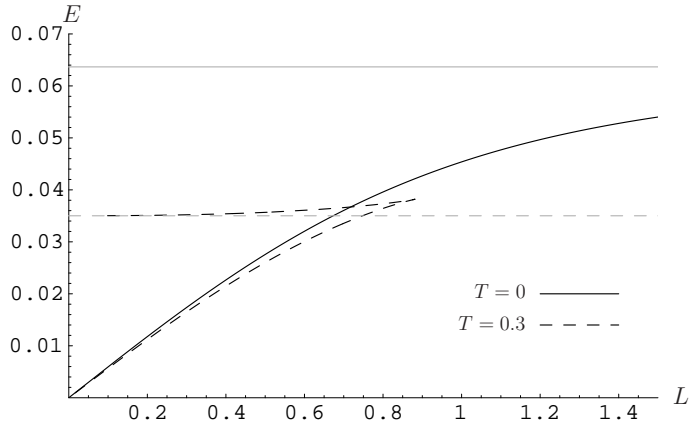}
\caption{The energy $E$
vs $L$, for $R=10(3/4 \pi)^{2/3}$, $r_{\textrm{\scriptsize max}}=200$ 
and $\alpha'=10^3 /GeV^2$. 
The solid and dashed curves represent 
the energy 
at $T=0$ and $T = 0.3$, respectively.
And the solid and dashed lines represented 
the energy of two straight strings at $T=0$ and $T=0.3$, respectively.}
\label{el46}
\end{center}
\end{figure}

\vspace{.5cm}
For the U-shape string configuration, we set the embedding coordinates as
follows:
\begin{eqnarray} 
\mathbf{X}_{||}=(\sigma, 0, 0) \ , \,\,\, \,\,\, 
T=\mbox{constant} \ , \,\,\, \,\,\, 
\Omega_4=\mbox{constant} \ .
\end{eqnarray}  
From the equation of motion for the coordinates $\mathbf{X}_{||}$,
we obtain the distance between quark and antiquark
\begin{eqnarray}
 L = \frac{2 R^{3/2} \sqrt{r_{min}^3-r_T^3}}{r_T^{1/2}}
\int_{r_{min}}^{r_{max}} \frac{dr}{\sqrt{(r^3-r_T^3)(r^3-r_{min}^3)}} \ ,
\end{eqnarray}
and, from the Hamiltonian,
we obtain the total energy of the U-shape string of inter-quark
separation $L$ 
\begin{eqnarray}
 E = \frac{1}{\pi \alpha'} 
\int_{r_{min}}^{r_{max}}  dr \sqrt{\frac{r^3-r_T^3}{r^3-r_{min}^3}} \ .
\label{energy46}
\end{eqnarray}
Fig.\ref{el46} shows the dependence of the energy $E$
on the distance $L$ at temperatures $T=0$ and $T \neq 0$. In this case,
both behaviors are equivalent to our gauge condensate model.

\section{Possible hadron spectrum at high temperature}

\subsection{Screening mass}

\begin{figure}[htbp]
\vspace{.3cm}
\begin{center}
  \includegraphics[width=7.5cm]{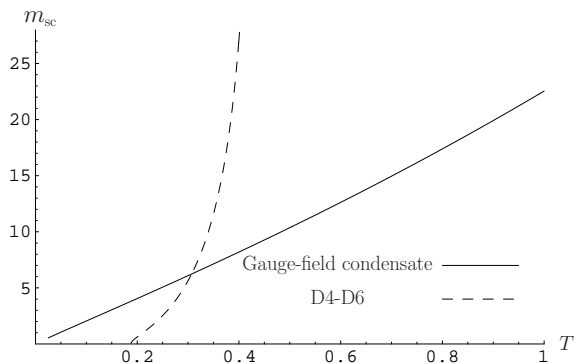}  
\caption{The temperature dependence of the screening mass:
the solid curve represents the result of the Gauge-field condensate model, 
while the dashed curve represents the result of the D4-D6 model.}
\label{screening}
\end{center}
\end{figure}

One of the basic characteristics of a plasma is the screening of
color electric fields.
In the paper~\cite{RTY}, 
the asymptotic behavior of the heavy quark potential at large distance
$L$ is given as follows;
\begin{eqnarray}
 V_{\mbox{\scriptsize{BN}}}(L,T) \approx 
-C_M \frac{e^{-m_{\textrm{\scriptsize sc}} L}}{L} + \cdots \ , 
\label{screeningmass}
\end{eqnarray}
where $m_{\textrm{\scriptsize sc}} \propto T$ is a screening mass.
In this section, we investigate the temperature dependence of the
screening mass in both the gauge-field condensate model and the D4-D6
model by comparing 
the quark-antiquark potential to Eq.~(\ref{screeningmass}).
The quark-antiquark potential in the gauge-field condensate model is given by 
\begin{eqnarray}
 V_{q \bar{q}}(L,T) = E(L,T) - 2 \tilde{m}_q \ . \label{hp}
\end{eqnarray}  
Comparing the two potentials~(\ref{hp}) and (\ref{screeningmass}) 
at distance $L^*$, we obtain the temperature dependence of the
screening mass as figure~\ref{screening}\footnote{We do not treat the
high temperature region because $L^*$ becomes too small to use the
results~(\ref{screeningmass}).}:
In the gauge-field condensate model, 
the screening mass is almost proportional to the  
temperature. This result coincides with the result given in Ref.~\cite{BN}.
While in the D4-D6 model,
the screening mass almost becomes zero in $T \sim 0.19$.
It is considered that the reason of this result is 
because quark and antiquark do not completely confine even at $T=0$.
As the temperature increases, 
the screening mass is almost proportional to the temperature 
until $T \sim 0.27$ and shapely increases when the temperature
exceeds $0.27$. This result is different 
from the one given in Ref.~\cite{BN}.

\subsection{Meson}

\begin{figure}[htbp]
\vspace{.3cm}
\begin{center}
  \includegraphics[width=9cm]{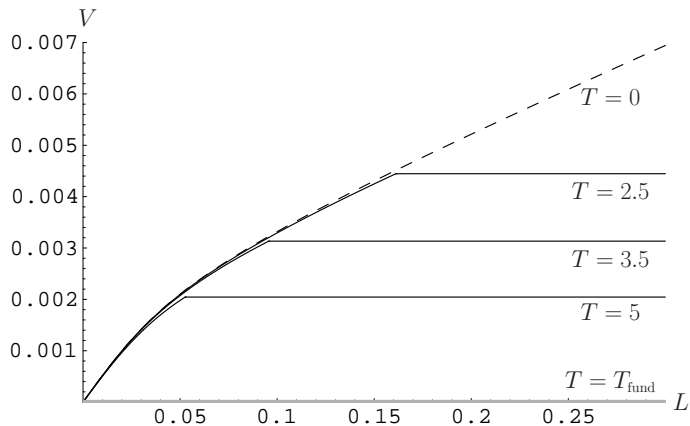}
\caption{The effective quark-antiquark potential $V$ for 
$R = 1/\sqrt{\pi} (\mbox{GeV}^{-1})$, 
$q=10^3 (\mbox{GeV}^{-4})$
 $r_{\mbox{max}}=10 (\mbox{GeV}^{-1})$ 
and $\alpha'=10^3 (\mbox{GeV}^{-2})$. 
The dashed curve represents the result for $T=0$ and the solid curves
 represent the results for $T=2.5$, $3.5$ and $5$, respectively.}
\label{mesonspectrum}
\end{center}
\end{figure}

In this section, we consider about the meson spectra. 
As mentioned in the end of section 3.1,
when $L \leq L^*$ the energy of the U-shape string becomes lower than 
the energy of the pair strings, while
when $L \geq L^*$ the result reverses. 
this result shows that when the distance between quark and antiquark is
close, these quarks are confined, while the distance becomes wide and
moreover exceeds $L^*$, deconfinement occurs and the energy
becomes two quark masses.
Therefore we can obtain effective potential like Fig.\ref{mesonspectrum}.  
According to Fig.\ref{mesonspectrum}, 
as temperature increase,
possibility of existence of meson spectra, 
which shall exist certainly at $T=0$, 
become lower and lower,
because the hight of the potential becomes shallower,
namely, the region in which mesons can exist becomes more narrow.
Especially, when the temperature exceeds 
$T_{\textrm{\scriptsize fund}}$,
the potential identically becomes zero.
This leads to 
the deconfinement.

We investigate this fact by a simple manner:
we solve the 3-dimensional Schr\"{o}dinger equation with this effective 
potential and investigate the bound states~\cite{Jackiw,Malik}.
As a result,
we could show that the bound states i.e. the meson spectra 
exist for $T \sim 0.043$ in the gauge-field condensate model.

Also, for comparison, utilizing the lattice data quoted from Fig.~3 
in reference~\cite{lattice},
we can investigate the bound states:
the heavy quark free energy converges finite value 
for long distance between quark. 
Therefore, we may regard the value as the mass of a pair of effective
quark and antiquark, which are not confining. 
Deforming the free energy so that we can regard the asymptotic value as
zero and using it as the potential of the Schr\"{o}dinger equation,
we show a result that the bound states exist for less than the critical
temperature $T_{\rm fund}$. 

\subsection{Baryon}

\begin{figure}[htbp]
\begin{center}
\voffset=15cm
  \includegraphics[width=9cm,height=7cm]{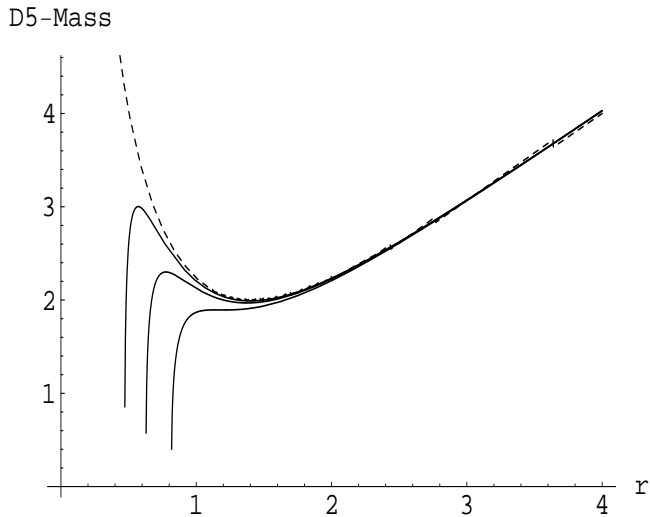} 
\caption{The D5-brane mass $M_{\rm D5}(r)$ as a function of $r$. 
Here we set $q=4$. $M_{\rm D5}(r)$ and $T$ are shown in units of 
$\tau_5\pi^3R^4$ and $R^{-2}$, respectively. 
The dashed curve shows the case of $T=0$, and 
three solid curves represent cases of $T=0.15,~0.2,~0.26$, 
respectively, from left to right. 
\label{D5mass} }
\end{center}
\end{figure}

It has been shown  
that baryons correspond to D5-branes wrapped around the 
compact manifold $M_5$ \cite{Gross,Witten}. As a typical case, 
here we take $S^5$ and investigate the qualitative property. 
The brane action of such a D5 probe is 
\beq
S_{\rm D5}= -\tau_5 \int d^6\xi e^{-\Phi} \sqrt{\cal G} \ ,
\label{D5-action}
\eeq
where $(\xi_i)=(X^0,X^5 \sim X^9)$, $\tau_5$ represents 
the tension of D5 brane, and 
${\cal G}=-{\rm det}({\cal G}_{i,j})$ for the induced metric 
${\cal G}_{ij}= \partial_{\xi^i} X^M\partial_{\xi^j} X^N G_{MN}$. 
The mass of the wrapped D5-brane is then 
\bea
M_{\rm D5}(r)=\tau_5 e^{-\Phi} \sqrt{\cal G} =
\tau_5 \pi^3 R^4 r f(r) e^{-\Phi/2} \; .
\eea
The mass $M_{\rm D5}$ thus defined depends on the position $r$ of the D5-brane. As for $T=0$,  $M_{\rm D5}$ has a simple form 
\beq
M_{\rm D5}(r)=\tau_5 \pi^3 R^4 r \sqrt{1+ {q \over r^4}} \; .
\eeq
The $M_{\rm D5}(r)$ diverges at both $r=0$ and $r=\infty$, and it 
has a global minimum $M_{\rm D5}(r_{\rm min})=\tau_5 \pi^3 R^4 (4q)^{1/4}$ 
at $r=r_{\rm min}=q^{1/4}$; see the dashed curve in Fig. \ref{D5mass}. 
The minimum value can be regarded as 
the baryon mass from the action principle. In the AdS limit, namely 
$q \to 0$, the baryon mass vanishes. Thus, the baryon mass is 
induced by finite $q$, i.e. by finite gauge-field condensate, in this model.

As for $T > 0$, $M_{\rm D5}(r)$ has a form 
\beq
M_{\rm D5}(r)=\tau_5 \pi^3 R^4 r 
\sqrt{
\left( 1+\frac{q}{r_T^4}\log({1\over 1-(r_T/r)^4}) \right) 
\left( 1- {r_T^4 \over r^4} \right)} \; .
\eeq
The $M_{\rm D5}(r)$ is real only for $r \ge r_T$. 
In the region, $M_{\rm D5}(r)$ is zero at $r=r_T$ and positive for $r > r_T$. 
Thus, even if a minimum exists at $r > r_T$, it is only a local minimum; 
see the solid curves in Fig. \ref{D5mass}. 
One could regard the local minimum as baryon, but the baryon is metastable when $T$ is finite.

Figure \ref{D5mass} shows the $r$ dependence of $M_{\rm D5}(r)$ 
for four values of $T$. 
When $T=0$, there exists a minimum 
at $r=q^{1/4}=1$, as mentioned above. The minimum 
becomes a local minimum for $T<0.26$, and eventually 
it disappears for $T>0.26$. 
Thus, baryon can survive as a metastable state 
for small $T$. 
Furthermore, we can see from Fig. \ref{D5mass} that the mass of 
the metastable baryon is almost 
independent of $T$.

\section{Summary}

A gauge theory with light flavor quarks is studied in a dual supergravity of 
the AdS background deformed by dilaton, which
induces the gauge-field condensate in the dual gauge theory. 
The high-temperature-phase background is constructed by making the 
AdS-Schwarzshild compactification. 
This background, at zero temperature limit, corresponds to 
a dual of the $\mathcal{N}=1$ 
supersymmetric gauge theory with the quark confinement \cite{GY}. 

Introducing the flavor quark by
embedding the D7 probe brane in this high-temperature
background, we found no spontaneous chiral symmetry breaking
in this case. Furthermore, through the analysis of the 
Wilson-Polyakov loop, we found
that the dynamical quark mass is not divergent and the quark-antiquark 
potential has a finite range. 
Thus, these properties are consistent with the one of high temperature
QCD phase. 

\vspace{.3cm}
It might be a new point that,
in this deconfinement phase, there exists still a phase transition 
at a temperature, $T_{\rm {fund}}$.
It is observed through the temperature dependence of the D7 brane energy 
and the vev of quark bilinear. 
In the gravity side, this transition is seen through the form of
the embedded D7 brane when its end-point meets with the horizon. 
This transition takes place for both 
cases with and without gauge-field condensate, and the
similar transition is also seen in 
other models \cite{KMMW2,Bab}. Hence, this
transition would be universal. 

\vspace{.3cm}
In the higher temperature phase ($T>T_{\rm {fund}}$), both
the dynamical quark mass and the potential between quark and antiquark
vanish. This implies that 
the phase is in a quark-gluon plasma. 
Meanwhile, 
in the lower temperature phase ($T<T_{\rm {fund}}$), 
the dynamical quark mass is finite and a short ranged interaction 
between quark and antiquark still remains. In consequence, 
quark bound states are possible. 
The bound states would be atom-like in the sense 
that each constituent can be separated, and these bound states disappear 
above the critical temperature $T_{\rm {fund}}$. 
Similar property can be seen for baryon, which is studied here by
D5 brane embedded in the high-temperature background. 

\vspace{.3cm}
As for the dynamical quark mass, its 
temperature dependence is compared with the
numerical results of full lattice QCD,  
and we found that our result is
qualitatively consistent with the lattice data.
 The temperature dependence of the screening mass is also investigated. 
We find that the screening mass increases linearly with temperature 
in our present model \cite{BN}. 
This is also consistent with the analysis given in real QCD.
The present analysis is valid in the large $N_{c}$ limit and D7 brane is 
treated as a probe. 
In this sense, our analysis is akin to the quenched approximation, namely
the quark-antiquark creation is not included in our analysis.
However this would not affect the qualitative property of our results, 
since the qualitative property of the analyses
is not changed so much between full lattice QCD and quenched lattice QCD \cite{KKLL}. 

\vspace{.3cm}
Thus, we could say that there are good correspondences between 
the deconfinement phase of the present background and that of real QCD. 
Finally we should comment on the unwanted modes which are not seen
in the real QCD. Here we start from IIB superstring theory, and the 
action for the D7 brane includes fermionic fields as the super-partner.
These fermionic modes do not correspond to any baryonic state in 
real QCD. So we expect that the masses of these state would be large
and decouple to our low energy theory. But this point is open at present
stage.

\vspace{.3cm}
\section*{Acknowledgments}

This work has been supported in part by the Grants-in-Aid for
Scientific Research (13135223, 14540271)
of the Ministry of Education, Science, Sports, and Culture of Japan.
This work is also in part by the
Grant-in-Aid for Scientific Research on Priority Areas
"Progress in elementary particle physics of the 21st century
through discoveries of Higgs boson and supersymmetry" (No.441).


\newpage

\begin{thebibliography}{99}

\bibitem{MGW}
J.~M.~Maldacena,
Adv.\ Theor.\ Math.\ Phys.\  {\bf 2}, 231 (1998) [hep-th/9711200].

S.~S.~Gubser, I.~R.~Klebanov and A.~M.~Polyakov,
Phys.\ Lett.\ B {\bf 428}, 105 (1998) [hep-th/9802109].

E.~Witten,
Adv.\ Theor.\ Math.\ Phys.\  {\bf 2}, 253 (1998) [hep-th/9802150].
A.M. Polyakov, Int. J. Mod. Phys. {\bf A14} (1999) 645,
        ({\tt hep-th/9809057}).
 
\bibitem{KK}
A.~Karch and E.~Katz, JHEP {\bf 0206}, 043(2003)  [hep-th/0205236].
\bibitem{KMMW}
M.~Kruczenski, D.~Mateos, R.C.~Myers and D.J.~Winters, JHEP {\bf 0307}, 049(2003)  [hep-th/0304032].
\bibitem{KMMW2}
M.~Kruczenski, D.~Mateos, R.C.~Myers and D.J.~Winters, [hep-th/0311270].
\bibitem{Bab}
J.~Babington, J.~Erdmenger, N.~Evans, Z.~Guralnik and I.~Kirsch, 
hep-th/0306018. 
\bibitem{ES} N.~Evans, and J.P.~Shock, Phys. Rev.{\bf D70}, 046002 (2004).
\bibitem{SS} T.~Sakai and J.~Sonnenschein, JHEP {\bf 0309},047 (2003).
\bibitem{NPR}
C.~Nunez, A.~Paredes and A.V.~Ramallo, JHEP {\bf 0312}, 024(2003)  
[hep-th/0311201].
\bibitem{GY} K. Ghoroku and M. Yahiro, 
Phys. Lett. {\bf B604}, 235 (2004).

\bibitem{sakaisugimoto}
T.~Sakai and S.~Sugimoto, hep-th/0412141.
\bibitem{BY} D. Bak and H. Yee, Phys.Rev. D71 (2005) 046003 [hep-th/0412170].
\bibitem{RTY} S.J. Rey, S. Theisen and J.T. Yee, Nucl. Phys. B527(1998)171.

\bibitem{Jackiw}
O.J.P.~Eboli, R.~Jackiw and  S.-Y.~Pi, Phys. Rev. {\bf D37}, 3557 (1988).

\bibitem{Malik}
G.P.~Malik, R.K.~Jha and V.S.~Varma, Astroph. J. {\bf 503}, 446 (1998);
Eur. Phys. J. {\bf A2} 105, (1998).

\bibitem{KS2}
A.~Kehagias and K.~Sfetsos, Phys.\ Lett.\ B {\bf 456}, 22(1999) 
[hep-th/9903109]. 
\bibitem{LT}
H.~Liu and A.A.~Tseytlin [hep-th/9903091].
\bibitem{GGP} G. W. Gibbons, M. B. Green and M. J. Perry, 
  Phys.Lett. B370 (1996) 37-44, [hep-th/9511080].

\bibitem{lattice} O.~Kaczmarek, S.~Ejiri, F.~Karsch, E.~Laermann, F.~Zantow,
 Prog. Theor. Phys. Suppl. {\bf 153}, 287 (2004); 
 P.~Petreczky, K.~Petrov, 
Phys. Rev. {\bf D70}, 054503 (2004).

\bibitem{Gross}
D.~J.~Gross and H. Ooguri, Phys.~Rev.{\bf D58}, 106002(1998) [hep-th/9805129]. 
\bibitem{Witten}
E.~Witten, JHEP{\bf~7}, 006(1998) [hep-th/9805112]. 


\bibitem{BN}
E.~Braaten and A.~Nieto, Phys.~Rev.~Lett. 74 (1995) 3530.



\bibitem{KKLL} O. Kaczmarek, S. Karsch, E. Laermann, M. L\"{u}tgemeier, Phys.
 Rev. {\bf D62} 034021 (2000).
\end{thebibliography}
\end{document}